\documentclass[sigconf,edbt,capitalise,nameinlink,noabbrev]{acmart-edbt2025}

\usepackage{siunitx} 
\usepackage[inline]{enumitem}
\usepackage{hyperref}
\usepackage{xcolor}
\usepackage[frozencache]{minted}
\usepackage{tcolorbox}
\usepackage{csquotes}
\usepackage{makecell}
\usepackage{fbox}
\usepackage{diagbox}
\usepackage{pbalance}
\setlist{nosep}

\Crefname{lstlisting}{Listing}{Listings}

\usepackage[nolist,nohyperlinks]{acronym}
\begin{acronym}
	\acro{AEX}{Asynchronous Enclave Exit}
	\acro{DB}{Database}
	\acro{DBMS}{Database Management System}
	\acro{EPC}{Enclave Page Cache}
	\acro{EDMM}{Enclave Dynamic Memory Management}
	\acro{GDPR}{General Data Protection Regulation}
	\acro{MEE}{Memory Encryption Engine}
	\acro{NUMA}{Non-Uniform Memory Access}
	\acro{OS}{Operating System}
	\acro{OLTP}{Online Transaction Processing}
	\acro{OLAP}{Online Analytical Processing}
	\acro{PRM}{Processor Reserved Memory}
	\acro{SEV}{Secure Encrypted Virtualization}
	\acro{SGX}{Software Guard Extensions}
        \acro{SSB}{Speculative Store Bypass}
	\acro{SW}{Software}
	\acro{TCB}{Trusted Computing Base}
	\acro{TCS}{Thread Control Structure}
	\acro{TDX}{Trust Domain Extensions}
	\acro{TEE}{Trusted Execution Environment}
	\acro{TLB}{Translation Lookaside Buffer}
	\acro{TLS}{Transport Layer Security}
	\acro{TME}{Total Memory Encryption}
	\acro{UCE}{UPI Crypto Engine}
	\acro{UPI}{Ultra Path Interconnect}
	\acro{VM}{Virtual Machine}
\end{acronym}

\newcommand{\sgxTwo}[0]{SGXv2}
\newcommand{\sgxOne}[0]{SGXv1}
\renewcommand{\paragraph}[1]{\textbf{#1}}

\DeclareSIUnit{\rows}{\text{\,rows}}

\setkeys{Gin}{width=\linewidth} 

\newtcolorbox{lessons}[1][]{
    width=\linewidth+4pt,
    top=0pt,bottom=0pt,left=0pt,right=0pt,boxsep=2pt,
    before skip=2pt,after skip=2pt,left skip=-3pt,
    boxrule=1pt,#1
}

\setcopyright{rightsretained}

\acmDOI{}

\acmISBN{978-3-89318-098-1}

\acmConference[EDBT 2025]{28th International Conference on Extending Database Technology (EDBT)}{25th March-28th March, 2025}{Barcelona, Spain}
\acmYear{2025}

\begin{document}
\title{Benchmarking Analytical Query Processing in Intel SGXv2}

\author{Adrian Lutsch}
\email{adrian.lutsch@cs.tu-darmstadt.de}
\orcid{0009-0008-7889-8152}
\affiliation{%
	\institution{Technical University of Darmstadt}
	\city{Darmstadt}
	\country{Germany}
}

\author{Muhammad El-Hindi}
\email{muhammad.el-hindi@cs.tu-darmstadt.de}
\orcid{0000-0001-5295-1316}
\affiliation{%
	\institution{Technical University of Darmstadt}
	\city{Darmstadt}
	\country{Germany}
}

\author{Matthias Heinrich}
\affiliation{%
	\institution{Technical University of Darmstadt}
	\city{Darmstadt}
	\country{Germany}
}

\author{Daniel Ritter}
\email{daniel.ritter@sap.de}
\orcid{0000-0001-6146-3365}
\affiliation{%
	\institution{SAP SE}
	\city{Waldorf}
	\country{Germany}
}

\author{Zsolt Istv\'an}
\orcid{0000-0002-4127-8573}
\email{zsolt.istvan@tu-darmstadt.de}
\affiliation{%
	\institution{Technical University of Darmstadt}
	\city{Darmstadt}
	\country{Germany}
}

\author{Carsten Binnig}
\orcid{0000-0002-2744-7836}
\email{carsten.binnig@cs.tu-darmstadt.de}
\affiliation{%
	\institution{Technical University of Darmstadt \& DFKI}
	\city{Darmstadt}
	\country{Germany}
}

\begin{abstract}
Trusted Execution Environments (TEEs), such as Intel's Software Guard Extensions (SGX), are increasingly being adopted to address trust and compliance issues in the public cloud.
Intel SGX's second generation (SGXv2) addresses many limitations of its predecessor (SGXv1), offering the potential for secure and efficient analytical cloud DBMSs.
We assess this potential and conduct the first in-depth evaluation study of analytical query processing algorithms inside SGXv2.
Our study reveals that, unlike SGXv1, state-of-the-art algorithms like radix joins and SIMD-based scans are a good starting point for achieving high-performance query processing inside SGXv2.
However, subtle hardware and software differences still influence code execution inside SGX enclaves and cause substantial overheads.
We investigate these differences and propose new optimizations to bring the performance inside enclaves on par with native code execution outside enclaves.
\end{abstract}

\maketitle

\section{Introduction}
\label{sec:intro}

\paragraph{The need for secure cloud DBMSs.} The last decade has seen a fundamental shift in where \acp{DBMS} run: public clouds have become the primary location where data is stored and processed.
While there are many benefits in running \acp{DBMS} in the cloud, such as scaling on demand, the cloud model puts a high stake on the cloud provider regarding the security of the data~\cite{pearsonPrivacySecurityTrust2010}.
Today, customers have to fully trust the cloud providers to keep the data safe and avoid any attacks that can result in data breaches or data corruption.
Sadly, there are well-publicized examples of cloud providers failing to provide these guarantees~\cite{whittakerDanishCloudHost2023,bonnetCloudAssetsBiggest2023}.

\paragraph{TEEs to the rescue?} 
Thus, all major cloud providers are moving to provide new offerings to prevent such problems.
A prominent technology deployed widely in the cloud is so-called \acp{TEE}.
A \ac{TEE} is a hardware-based solution that shields a process from a potential attacker and has been successfully used to build secure DBMSs in the cloud~\cite{antonopoulosAzureSQLDatabase2020}.
On a high level, TEEs provide two primary protection guarantees.
First, they provide integrity, i.e., ensuring that software or hardware attacks cannot manipulate code and data without being detected.
Second, they guarantee confidentiality, i.e., code and data are encrypted inside a TEE and can not be accessed outside an enclave.

\paragraph{Security does not come for free.}
One of the first broadly available TEE technologies was Intel's \ac{SGX}.
SGX extends Intel CPUs with instructions and hardware components that enable ``secure enclaves'',  protecting processes against malicious administrators, operating systems, and hypervisors. However, being targeted for mobile and consumer devices, the first generation of SGX (SGXv1) had severe hardware limitations when used for DBMSs. In particular, memory access had a high overhead due to encryption and integrity checks, and the protected memory region that enclaves could access efficiently was only \SI{256}{\mega\byte}, leading to high overheads when the data sizes exceeded that limit.
As a result, \acp{DBMS} deployed on SGXv1 typically faced orders of magnitude slowdowns~\cite{maliszewskiWhatPriceJoining2021}, making the first generation of SGX unpractical for data-intensive systems ~\cite{el-hindiBenchmarkingSecondGeneration2022, maliszewskiWhatPriceJoining2021, priebeEnclaveDBSecureDatabase2018}.

\paragraph{Recent advances of SGX lift limitations.}
With the Intel Ice Lake architecture, Intel SGX became available on multi-socket server hardware~\cite{johnsonSupportingIntelSGX2021}. This second generation of Intel SGX (SGXv2) uses redesigned hardware to achieve isolation and confidentiality guarantees. Most importantly, the new generation relieves the memory limitation issue by allowing enclaves to access up to \SI{512}{\giga\byte} encrypted memory per socket~\cite{johnsonSupportingIntelSGX2021,el-hindiBenchmarkingSecondGeneration2022}. Additionally, integrity checks have been streamlined, and enclave processes can leverage the newly added multi-socket support.
After releasing the second generation of SGX, Intel discontinued the first generation.

\begin{figure}[tp]
    \centering
    \includegraphics{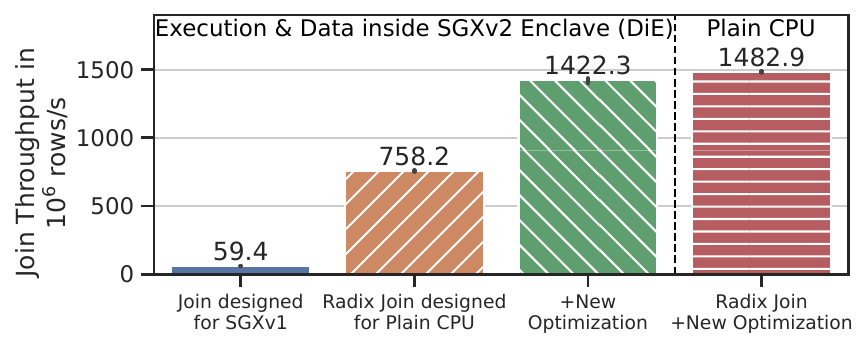}
    \caption{Performance of joining a \qty{100}{\mega\byte} (hash) and a \qty{400}{\mega\byte} (probe) table inside an SGXv2 enclave. The join designed for SGXv1 does not achieve competitive performance (blue). A state-of-the-art radix join is a better starting point (orange), and with our optimization (green), its  performance is similar to outside the enclave (red).}
  \label{fig:intro_plot}
\end{figure}

\paragraph{The need for a performance study of SGXv2.}
While SGXv2 promises many benefits over SGXv1, the impact of integrating SGXv2 in the design of secure DBMSs is not yet well understood. 
For example, it has not yet been explored whether SGXv2 can meet the demands of high-throughput analytical query execution operators.
Hence, this paper provides the first in-depth study of running query execution operators in SGXv2.
Our study is motivated by the observation that previous design principles to improve performance in SGXv1 by optimizing for the limited enclave memory as the main bottleneck are not adequate anymore. 
Instead, state-of-the-art data processing algorithms that target server-grade hardware and include optimizations like cache consciousness are a better starting point, as shown in Figure~\ref{fig:intro_plot}.

In this initial experiment, we compare a join designed for SGXv1 \cite{maliszewskiCrackingLikeJoinTrusted2023} to a cache-optimized radix join (both executed on SGXv2 hardware).
The results in \cref{fig:intro_plot} illustrate that the join designed for SGXv1 (blue bar) achieves a much lower performance compared to a state-of-the-art join implementation (orange bar).
This performance difference arises because the SGXv1 optimized join prioritizes avoiding enclave paging -- a concern irrelevant in SGXv2 -- at the expense of not fully utilizing all available cores on the server-grade SGXv2 hardware.
However, it also becomes clear that the radix join inside the enclave does not match the performance when executed outside of the enclave (red bar).
As we uncover in this study, this performance gap results from different micro-architectural behaviors of running code inside and outside an SGXv2 enclave.
To address these micro-architectural differences, we discuss new optimizations allowing DBMSs to achieve almost native performance as exemplified by the SGXv2-optimized join in Figure \ref{fig:intro_plot} (green bar).

\paragraph{Focus on analytical query processing.} 
Rich related work has underscored that OLAP DBMSs can only achieve high performance and efficiency if the underlying CPU micro-architecture is taken into account~\cite{polychroniouRethinkingSIMDVectorization2015,balkesenMainmemoryHashJoins2013,kerstenEverythingYouAlways2018}. Given the importance of micro-architectural effects in the context of OLAP and the under-explored performance characteristics of SGXv2, our study focuses on in-memory OLAP workloads. 
We implemented state-of-the-art (micro-architecture-aware) joins~\cite{balkesenMainmemoryHashJoins2013} and column scans~\cite{willhalmSIMDscanUltraFast2009}, query execution operators that are at the core of all OLAP databases. This allows us to study their performance characteristics, uncover performance pitfalls, and provide suggestions for designing efficient algorithms for SGXv2.

\paragraph{Contribution and main findings.}
To summarize, in this paper, we present the results of the first in-depth performance study of key OLAP query execution operators in SGXv2 enclaves.
Our study makes the following core contributions:
\begin{enumerate}
\item We show that state-of-the-art main memory and cache-optimized algorithms are a better starting point for \sgxTwo{} than algorithms optimized for \sgxOne{}: I.e., previously suggested SGX-optimized joins are no longer required, and throughput-optimized
scan algorithms work at nearly equal performance out of the box.
\item We study the hardware and software overheads of state-of-the-art cache-optimized database algorithms in \sgxTwo{} and uncover unknown overheads that are caused by a side channel mitigation enabled inside the enclave but not outside. 
Based on a radix join, we show how existing algorithms can be optimized to circumvent this previously unknown slowdown and match the throughput of join processing outside of \sgxTwo{} enclaves.
\item Finally, we show that these results can be generalized to the execution of query plans, resulting in performance almost on par with native execution outside the enclave. I.e., the additional security of \sgxTwo{} enclaves comes with only minor performance costs for query execution.
\end{enumerate}

Being a performance study, this work is not concerned with the security properties of Intel SGXv2. Thus, we do not investigate specialized data structures and algorithms meant to prevent information extraction via side channels, such as algorithms with oblivious memory access patterns. Instead, we focus on the performance costs of the security technology and regard a detailed analysis of its security guarantees and weaknesses as future work.

\paragraph{Outline.}
The rest of this paper is structured as follows. First, \cref{sec:background,sec:overview} give the necessary background about Intel \sgxTwo{} and our benchmark setup. Afterward, in \cref{sec:join,sec:scans,sec:full-queries}, we evaluate the performance of key OLAP algorithms, join and scan, in-depth and then study their composition in query plans. Finally, \cref{sec:summary,sec:related,sec:conclusion} discuss a potential performance model, present related work, and conclude this performance study.

\begin{figure}
\center
    \centering
    \includegraphics[width=1\columnwidth]{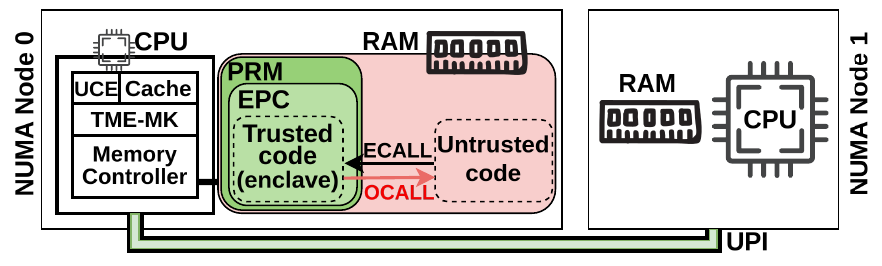}
    \caption{Intel SGX implements enclaves via a protected memory region in RAM, called \acf{PRM}. Data and code of enclaves are stored in encrypted memory pages inside the \acf{EPC}. They are decrypted when loaded into the cache. The \acf{UCE} encrypts enclave UPI traffic.}
    \label{fig:sgx-architecture}
\end{figure}

\section{Intel \sgxTwo{} Background}
\label{sec:background}

The new server-grade generation of Intel \ac{SGX} introduced with the Intel Ice Lake architecture \cite{johnsonSupportingIntelSGX2021} lifts several limitations of the first generation that led to high overheads in terms of performance.
In this section, we review the basics of Intel's \ac{SGX} technology and discuss the most important changes of SGXv2.

\paragraph{Integrity and confidentiality in SGX.}
Intel \ac{SGX} protects the integrity of user code by shielding it even from privileged entities like the \ac{OS} or the hypervisor.
On a high level, this guarantee is achieved by creating a protected memory region in RAM, called \acf{PRM}, which can only be accessed via special CPU instructions~\cite{mckeenInnovativeInstructionsSoftware2013, costanIntelSGXExplained2016}.
As shown in \Cref{fig:sgx-architecture}, inside this protected memory region, SGX maintains the \acf{EPC} (light green area) to enforce enclave isolation.
The \ac{EPC} stores the trusted code and data of enclaves within encrypted \qty{4}{\kilo\byte} memory pages.
These pages are only decrypted when loaded into the CPU cache for processing~\cite{mckeenInnovativeInstructionsSoftware2013,costanIntelSGXExplained2016}.
Intel \ac{SGX} guarantees that only trusted code from within the same enclave has access to the \ac{EPC} pages of that enclave by adding security checks to the address translation.
Importantly, code running in the untrusted memory region outside the \ac{PRM} (including the \ac{OS}) is prevented from reading and modifying these pages.

\paragraph{Major differences in \sgxTwo{}.}
While the capacity limitations of the PRM made Intel's \sgxOne{} impractical for data-intensive applications such as DBMSs~\cite{priebeEnclaveDBSecureDatabase2018,maliszewskiWhatPriceJoining2021,el-hindiBenchmarkingSecondGeneration2022}, the new \sgxTwo{} design supports up to \qty{512}{\giga\byte} of \ac{PRM} per socket, which allows \acp{DBMS} to hold large data sets fully in the \ac{EPC} and avoids expensive enclave paging. This was achieved by replacing the previously used SGX Memory Encryption Engine with the new Total Memory Encryption -- Multi-Key (TME-MK)~\cite{johnsonSupportingIntelSGX2021}. In addition to changing the encryption hardware, SGXv2 replaced the integrity protection and freshness tree and the associated checks when loading encrypted enclave data into the cache with specialized bits in ECC memory~\cite{johnsonSupportingIntelSGX2021}.
Finally, enclaves can now scale across multiple CPU sockets, increasing the number of CPU cores and the amount of memory available even further~\cite{johnsonSupportingIntelSGX2021}.
To access \ac{EPC} pages on a remote socket, \sgxTwo{} introduces an additional \acf{UCE} that encrypts data before transferring it over \ac{UPI}~\cite{johnsonSupportingIntelSGX2021} (cf.~\Cref{fig:sgx-architecture}).

\paragraph{Implications of \sgxTwo{} for DBMSs.}
Although our previous work~\cite{el-hindiBenchmarkingSecondGeneration2022} indicates that with the second generation, Intel \ac{SGX} has become a viable option for OLTP workloads, many important \sgxTwo{} characteristics in the context of OLAP remain unexplored.
For example, it is unclear if the new memory encryption hardware can keep up with the high throughput demands of optimized column scan algorithms. Furthermore, while we studied the latency of random cross-\acs{NUMA} memory accesses in the context of OLTP~\cite{el-hindiBenchmarkingSecondGeneration2022}, we did not analyze the effects on throughput and query execution operators like joins, which is essential for analytical query processing.
Throughput-optimized OLAP algorithms using multiple threads have only been studied in the context of \sgxOne{}~\cite{maliszewskiWhatPriceJoining2021,maliszewskiCrackingLikeJoinTrusted2023}.
However, with the hardware changes of \sgxTwo{} 
mentioned above, it remains unclear if these findings of OLAP processing on \sgxOne{} 
generalize to \sgxTwo{} and how the hardware characteristics of \sgxTwo{} affect query execution performance. We address these questions in this paper.

\section{Benchmark Overview}
\label{sec:overview}

In the following, we give an overview of the benchmark settings, the framework, the used hardware, and the scope of the study. The input data for the algorithms is described at the beginning of the corresponding evaluations in \cref{sec:join,sec:scans,sec:full-queries}.

\paragraph{Benchmarking settings.}
The main idea of our evaluation study is to analyze the characteristics of \sgxTwo{} by comparing the performance of join and scan algorithms, both when executed natively on the CPU without security extensions and in an enclave.
Therefore, we compare two settings:
\begin{enumerate}
\item \emph{Plain CPU.} Traditional query processing baseline where the code is natively deployed on the CPU. This mode provides no security protections but also does not come with any additional overheads for computation and memory accesses. Data is always stored in untrusted memory in this setting.
\item \emph{SGX Data in Enclave (DiE).} Code and data are deployed within an enclave for processing. Since data resides in the \ac{EPC}, it undergoes (transparent) decryption when loaded into CPU caches and encryption when writing data back to memory.
\end{enumerate}
Additionally, we leverage the fact that Intel SGX enclaves can access the unprotected memory of their host process. This mechanism is necessary for communication with the enclave and can be used to trade security for performance, e.g., by storing non-sensitive data outside the enclave. It results in a third setting:
\begin{enumerate}[resume]
\item \emph{SGX Data outside Enclave (DoE).} Data is stored in untrusted (non-protected) memory, while code is processed within the enclave. This setting eliminates memory encryption/\allowbreak decryption overheads and allows us to distinguish between slowdowns caused by memory encryption and slowdowns caused by code execution within an enclave.
\end{enumerate}
By comparing the behavior of joins and scans in these settings, we seek to identify computation and memory access patterns that exhibit different throughput or latency behaviors, enabling us to understand and optimize for the characteristics of SGXv2.

\paragraph{Benchmarking framework.}
We implement all our query processing operators either based on published best practices in the OLAP literature (e.g.,~\cite{willhalmSIMDscanUltraFast2009,polychroniouRethinkingSIMDVectorization2015} for column scans) or based on existing benchmarks such as TEEBench~\cite{maliszewskiWhatPriceJoining2021}.
Moreover, to reveal the root causes of performance bottlenecks, we use self-implemented micro-benchmarks.
All benchmarking code is written in C/C++ and compiled with GCC version 12.3 using the optimization flags \texttt{-O3 -march=native} to ensure the highest optimization for our target architecture.
To implement code running inside the SGXv2 enclave, we use the (default) SGX SDK provided by Intel in version 2.24.
For measuring execution times, we rely on the RDTSCP instruction\footnote{
Stands for \emph{Read Time-Stamp Counter and Processor ID~\cite{intelcorporationIntel64IA322023a}}}
since it is the only available method to measure execution times (as CPU cycles) with high precision in both CPU modes.
If not otherwise stated, measurements are started after all required data for an operation has been allocated and initialized.
This approach allows us to minimize the impact of, e.g., context switches and measure only the execution performance of the actual query processing algorithms.
Similarly, our benchmarks only use data sizes that fit completely into the \ac{EPC} to prevent the paging costs from dominating the measurements.
We execute all experiments ten times and report the arithmetic mean and standard deviation.

\begin{table}
    \centering
    \caption{Hardware used for our benchmarks.}
    \small
    \begin{tabular}{ |c|c|c| }
        \hline
        Processor Name        & Intel Xeon Gold 6326 \\
        Sockets               & 2                    \\
        Cores per socket      & 16                   \\
        Threads per socket    & 16 (HT disabled)     \\
        Base Frequency        & 2.9 GHz              \\
        L1d Cache (per core)  & 48 KB                \\
        L1i Cache (per core)  & 32 KB                \\
        L2 Cache (per core)   & 1.25 MB              \\
        L3 Cache (per socket) & 24 MB                \\
        Microcode version     & 20240312 \\
        Memory Channels (per socket) & 8 \\
        Memory                & 16 * 32 GB \\
        Memory Speed and Latency & DDR4 3200 22-22-22 \\
        Memory Type & RDIMMs with ECC \\
        EPC size (per socket) & 64 GB \\
        \hline
    \end{tabular}
    \label{tab:hardware}
\end{table}

\paragraph{Benchmarking hardware.}
For all experiments, we use a dual-socket server featuring 3rd Generation Intel Xeon Scalable, \sgxTwo{}-capable processors with 16 cores.
The system is equipped with \qty{512}{\giga\byte} ($256$ per socket) main memory distributed over 16 DIMMs that populate all memory channels of both sockets (see \cref{tab:hardware} for more detailed hardware characteristics).
Our server runs Ubuntu 22.04.4 with kernel version 6.5 and uses the latest processor microcode (\texttt{20240312/0d0003d1}).
Following security guidelines for SGX, we disabled Hyper Threading on the CPUs.
To prevent noise caused by CPU frequency changes, we disabled Turbo Boost, 
changed the maximum CPU frequency to the base frequency (\qty{2.9}{\giga\hertz}), and enabled the performance governor to keep the CPU cores consistently on this fixed frequency.
To prevent NUMA effects from influencing experiments, we pin execution threads to one NUMA node. On our trusted operating system, this is possible by pinning threads outside of the enclave with \texttt{numactl} or \texttt{pthreads} since the threads stay pinned to their core upon entering the enclave.

\paragraph{Study overview.}
Our study is split into three main parts.
First, we analyze the performance effects of SGXv2 for joins. Secondly, we examine the throughput of multi-threaded column scans employing SIMD instructions. 
Finally, we study the performance of both operators in query plans.

\clearpage
\section{Join Algorithms in \sgxTwo{}}
\label{sec:join}

Joins are performance-critical operators in analytical databases because they involve processing large amounts of data.
Hence, they have recently been studied also in the context of the first generation Intel SGX hardware~\cite{maliszewskiWhatPriceJoining2021}.

\textbf{Importance of the analysis.}
As shown in the introduction, simply adopting algorithms optimized for SGXv1 does not result in high performance.
Further, despite offering enough enclave memory, the new SGXv2 hardware and security mechanisms still influence the performance of state-of-the-art join algorithms like the radix join.
In this section, we analyze the root cause of these slowdowns in-depth by studying different classes of join algorithms with varying memory access patterns.

\textbf{Join algorithms.} For the investigation, we have built our own benchmark suite based on TEE\-Bench~\cite{maliszewskiWhatPriceJoining2021}, a collection of parallel join algorithm implementations for benchmarking \sgxOne, and optimized the joins for SGXv2.
To gain an overview of how the SGXv2 hardware affects the performance of standard join algorithms, we use the following implementations:

\begin{enumerate}
    \item \emph{Hash join (PHT).} The \emph{Parallel Hash Table Join}~\cite{blanasDesignEvaluationMain2011} uses multiple threads to create a shared hash table from the smaller join input table. Afterward, the threads iterate over partitions of the larger input table, probing the hash table. It uses a classical bucket chaining hash table and enables parallel writes to the hash table by latching the buckets.
    \item \emph{Radix join (RHO).} The \emph{Radix Hash Optimized}~\cite{manegoldOptimizingMainmemoryJoin2002} join first partitions both input tables into cache-sized partitions by the least significant bits of their join key. To join the partitions, it employs an optimized hash table design, which achieved the best performance in previous evaluations~\cite{balkesenMainmemoryHashJoins2013,maliszewskiWhatPriceJoining2021} (implementation from~\cite{balkesenMainmemoryHashJoins2013}). The implementation studied here uses a two-phase parallel hash partitioning method similar to the method described in~\cite{kimSortVsHash2009}.
    \item \emph{Sort merge join (MWAY).} Sort merge joins first sort both input tables and then scan the sorted tables 
    for matching rows in one pass. We added the implementation of the Multi-Way Sort Merge Join (MWAY)~\cite{kimSortVsHash2009} from TEEBench to our benchmark suite.
    \item \emph{Index nested loop join (INL).} 
    The \emph{Index Nested Loop Join}~\cite{maliszewskiWhatPriceJoining2021} (INL) in our evaluation uses an existing B-Tree index to find matching tuples for every tuple in the outer table.
\end{enumerate}

In addition to these join algorithms, which are not optimized for SGX, we also investigate CrkJoin~\cite{maliszewskiCrackingLikeJoinTrusted2023}. CrkJoin is a partitioned hash join designed with the main bottlenecks of \sgxOne{} in mind: EPC paging and random main memory accesses. It performs in-place radix partitioning without random memory accesses by iteratively sorting input tables into partitions. The sort happens one bit at a time. Two pointers are moved from the start and end of the table towards the middle until they meet. Tuples with keys in the wrong order are swapped. The sort starts single-threaded and the number of sorting threads is doubled after each bit until the number of hardware threads is reached. After partitioning, CrkJoin uses the same in-cache join method as RHO~\cite{maliszewskiCrackingLikeJoinTrusted2023}. We optimized CrkJoin for our experiment hardware by configuring the available L2 cache size, as mentioned by the authors.

We do not study specialized join implementations that hide memory access patterns like the oblivious hash join of ObliDB~\cite{eskandarianObliDBObliviousQuery2019} or the oblivious sort-merge join from Opaque \cite{zhengOpaqueObliviousEncrypted2017}. We excluded them because they include algorithmic overheads that make them much slower than traditional join algorithms \cite{maliszewskiWhatPriceJoining2021} and might hide performance overheads caused by architectural effects of SGXv2.

\textbf{Join data.}\label{par:join-data} The input tables consist of rows with a 32-bit key (as join columns) and a 32-bit value (as tuple payload). All joins are foreign key joins, and keys follow a random uniform distribution.
Similar to previous studies~\cite{maliszewskiWhatPriceJoining2021,maliszewskiCrackingLikeJoinTrusted2023,blanasDesignEvaluationMain2011,kimSortVsHash2009,schuhExperimentalComparisonThirteen2016}, we do not materialize join results in most of our join benchmarks to prevent potential side-effects caused by expensive memory allocations. We look at the effect of result materialization and memory allocation separately in \cref{sec:join-sdk} and in full queries in \cref{sec:full-queries}. The experiments in this section join a \qty{100}{\mega\byte} and a \qty{400}{\mega\byte} table if not mentioned otherwise. This equals the cache-exceed setting in the TEEBench paper~\cite{maliszewskiWhatPriceJoining2021} on join performance in SGXv1 and is similar to average join sizes in TPC-H at scale factor 100. 
Since our goal is not a comprehensive comparison of join algorithms among each other but understanding the performance effect of SGXv2 enclaves on join execution, we do not investigate varying payload sizes, data distributions, or other data characteristics. In particular, we do not investigate data skew since it causes non-SGX-specific effects, such as stragglers during parallel execution. Instead, we focus on those parameters that uncover interesting performance and hardware effects of SGXv2.

\begin{figure}
    \centering
    \includegraphics{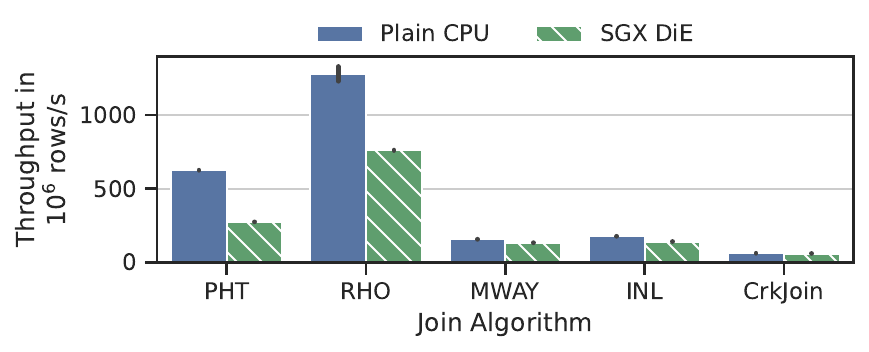}
    \caption{Overview of join algorithm throughput for 5 different joins executed using 16 threads to join a \qty{100}{\mega\byte} and a \qty{400}{\mega\byte} table on SGXv2 hardware. The SGXv1-optimized CrkJoin is the slowest join in this comparison. The hash joins have the highest slowdowns.}
    \label{fig:join-overview}
\end{figure}

\paragraph{Initial results.} \Cref{fig:join-overview} gives an overview of the throughput of the join implementations in our benchmark. Throughput is expressed as the sum of input cardinalities (numbers of rows) divided by the join execution time. All 16 hardware threads on one socket are used for execution. We compare the performance of the same join implementation running inside an SGX enclave with all inputs, intermediate data structures, and outputs stored inside the enclave (SGX Data in Enclave/DiE) with a plain CPU baseline that runs the join without an enclave.

This experiment shows several interesting insights:
Firstly, the performance of all join algorithms is lower when executed inside an SGXv2 enclave.
Secondly, the reduction in throughput of all these joins when executed inside the enclave varies considerably between join types. The hash joins PHT and RHO have a much higher performance overhead than MWAY, INL, and CrkJoin.
Finally, CrkJoin is the slowest join in our overview, reaching only \qty{60}{\mega\rows\per\second}. When executed inside the enclave, all other join algorithms perform better than CrkJoin, with speedups between $3\times$ for INL and $12\times$ for RHO. The measured performance of CrkJoin aligns with what the authors reported for SGXv1 hardware~\cite{maliszewskiCrackingLikeJoinTrusted2023}, and in our experiments, performance was similar independently of table sizes and skew. Thus, we derive that CrkJoin does not profit from the less restricting SGXv2 hardware. This can be attributed to its sort-based partitioning method, which starts with a single thread sorting on the first bit and then creates exponentially more threads for the following bits until all cores are used. In contrast, the other join implementations always use all available cores in parallel. 
CrkJoin's partitioning method is only competitive with other partitioning methods if memory access is severely bottlenecked by the EPC paging, as is the case in SGXv1.
However, on the new SGXv2 hardware, EPC paging is no longer a major performance limitation, and parallelism is important to achieve competitive throughput. Thus, CrkJoin is slower than the other joins in SGXv2.

\begin{lessons}
\paragraph{Lessons learned.} The main memory and cache-optimized join algorithms perform better inside SGXv2 than the SGXv1-optimized CrkJoin due to changed hardware characteristics, but they exhibit significant overheads.
\end{lessons}

\paragraph{Root causes of overheads.}
As we will show in the rest of this section, the slowdowns visible in the overview can be attributed to factors originating from the SGX security mechanisms on a hardware level. Additionally, there are other important performance factors that are rooted not purely in hardware but also in the software (e.g., the SGX SDK). We first summarize these factors below and present more detail in \cref{sec:join-random-access,sec:join-pipelining,sec:join-numa,sec:join-sdk}:

\begin{enumerate}
    \item \emph{Hardware-only effects.} Two hardware factors cause the slowdown of the hash joins in the overview. The first (cf. \cref{sec:join-random-access}) is the more expensive random main memory access inside the enclave. Optimizing to mitigate this known effect is more important in SGXv2 since EPC paging is no longer the limiting factor. Additionally, we uncover a previously unknown overhead that does not result from SGX-specific memory encryption and security checks but from a side channel mitigation that is always enabled in SGX enclaves. This issue is investigated in \cref{sec:join-pipelining}, where we also demonstrate how manual loop unrolling and instruction reordering can alleviate it.
    \item \emph{Mixed effects.} Other important performance effects result from an interplay of SGX software (i.e., the SGX SDK and the OS) and the SGX hardware. Firstly, while the support for \ac{NUMA} in SGXv2 enables the usage of more cores in joins, \cref{sec:join-numa} reveals that the unavailability of NUMA-awareness in SGX enclaves causes slowdowns because cross-NUMA traffic for joins can not be avoided. Secondly, \cref{sec:join-sdk} demonstrates how thread synchronization and memory management for SGXv2 can cause significant slowdowns if not handled carefully.
\end{enumerate}

\subsection{Overhead of Random Accesses}
\label{sec:join-random-access}

As mentioned before, random main memory access is a performance problem known from previous studies on SGXv1 \cite{maliszewskiWhatPriceJoining2021,maliszewskiCrackingLikeJoinTrusted2023} and our own evaluation on OLTP workloads in SGXv2 \cite{el-hindiBenchmarkingSecondGeneration2022}. 
In the following, we investigate the performance effects of slower random access on join algorithms. 
We use the Parallel Hash Table (PHT) join as an example because it is not optimized to reduce cache misses. In our investigation, we vary the size of the smaller input table (build side) from \qty{1}{\mega\byte} to \qty{100}{\mega\byte} and measure the join throughput in enclave relative to the throughput outside of the enclave. The probe table size is fixed at \qty{1}{\giga\byte} and only a single join thread is used to prevent parallelization effects from influencing the measurements. To investigate the correlation between cache misses and performance, we measure the number of Last Level Cache (LLC) misses during the join and report this number divided by the sum of the input table cardinalities as \emph{LLC misses per row}.

\begin{figure}
    \centering
    \includegraphics{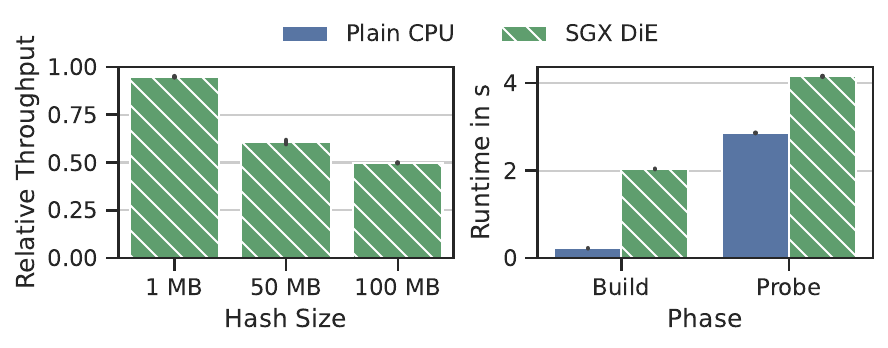}
    \caption{Left: Throughput of a single-threaded hash join with data and execution inside an SGXv2 enclave (DiE) relative to plain CPU.
    Join performance with large hash tables suffers from random access overhead. 
    Right: Comparison of join phase runtimes at \qty{100}{\mega\byte} hash size. The slowdown of the build phase inside the enclave is significant.}
    \label{fig:random-memory-access-join}
\end{figure}

The experiment results are depicted on the left side of \cref{fig:random-memory-access-join}. The first bar shows that for a small table size of \qty{1}{\mega\byte}, which fits into the cache of the tested CPU and causes fewer than 0.1 LLC misses per row, the join throughput inside the enclave is \qty{95}{\percent} of the throughput outside the enclave. When increasing the size of the smaller table to \qty{50}{\mega\byte} and \qty{100}{\mega\byte}, which is 4 times larger than L3 cache, the number of cache misses increases to 1.75 and 1.9 LLC misses per row and the relative performance decreases to \qty{62}{\percent} and \qty{51}{\percent} respectively. Thus, the relative performance of the join correlates with the frequency of cache misses.

The next interesting question is which of the two join phases (building the hash table and probing it) affects performance most. Thus, we break down the hash join runtime into phases in \cref{fig:random-memory-access-join}, right part. It reveals that the random-write-heavy build phase suffers a considerably higher performance reduction than the join phase. This raises the question if memory writes in SGXv2 have higher overheads than reads.

\paragraph{Random main memory access micro-benchmark.}
To answer if slower writes cause the higher overhead measured for the hash table build phase, the following micro-benchmark compares reading and writing 8-byte integers at random positions inside an array. The positions are determined by a linear congruential generator. We measure the throughput and vary the array size.

\begin{figure}
    \centering
    \includegraphics{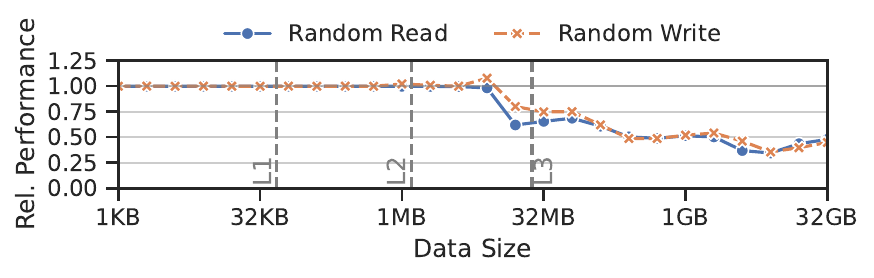}
    \caption{Performance of random memory reads and writes in an SGX enclave relative to plain CPU. In the cache, random access performance is equal. Random accesses to main memory are significantly slower in SGXv2.}
    \label{fig:pmbw-random}
\end{figure}

The results are depicted in \cref{fig:pmbw-random}. 
We derive three main insights:
\begin{enumerate*}
    \item If the data is cache-resident, random memory reads and writes have no performance penalty inside SGX (as expected).
    \item When increasing array sizes to larger than the cache sizes, the relative performance of random reads and writes decreases to similar degrees. Therefore, the bigger slowdown of the join's build phase cannot be explained by a performance difference between random reads and writes.
    \item The performance decreases are significant. We see nearly 3x  higher in-enclave latencies for the \qty{8}{\giga\byte} array size and a doubling in latencies at \qty{256}{\mega\byte}, which is the size of the hash table created in the join benchmark above.
\end{enumerate*}

We attribute the observed slowdown to two features of the SGXv2 hardware: 
\begin{enumerate*}
    \item Since enclave memory is transparently encrypted, data must be decrypted when read from memory into the cache and encrypted when written to memory. According to measurements by Intel for TME-MK, this adds 11 ns of latency to last level cache misses \cite{bronleeweRuntimeEncryptionMemory2022}.
    \item Most of the security guarantees of Intel SGX are enforced by adding checks to address translation \cite{costanIntelSGXExplained2016}. This increases the cost of TLB misses. Thus, the constant TLB misses caused by random accesses over large memory areas are extra costly to resolve in SGX enclaves, leading to significantly higher random access latencies and reducing throughput.
\end{enumerate*}

\begin{lessons}
\textbf{Lessons learned.} Random main memory access in SGXv2 enclaves causes high performance overheads that lead to a significant slowdown of algorithms and data structures dependent on them, such as the PHT and INL join, grouping, hash tables, and trees. Our micro-benchmarks show up to three times worse random main memory access performance in SGXv2. When data fits in the cache, there is no overhead caused by slower random memory access. Thus, there is a strong incentive to employ techniques that either keep data cache-resident for processing or hide the latency of cache misses.
\end{lessons}
One open question is why the hash table creation (build phase) is 9x slower inside the enclave, although the random memory access micro-benchmark only explains a 3x slowdown. As discussed in the next section, the underlying reason for this discrepancy and the slowdown of the RHO join is the same.

\subsection{Overhead of Side Channel Mitigation}
\label{sec:join-pipelining}

The RHO join does not suffer from random main memory access overheads because of its
partitioning. However, the overview in \cref{fig:join-overview} still reveals performance reductions of more than \qty{30}{\percent}.

\begin{figure}
    \centering
    \includegraphics{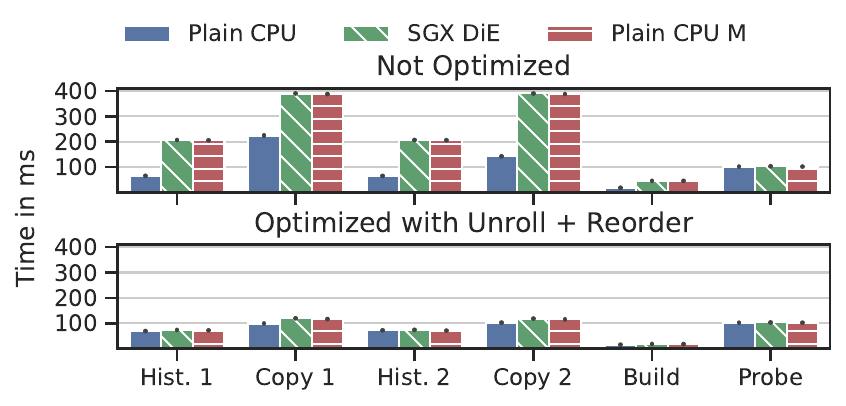}
    \caption{Runtime breakdown for the phases in a single-threaded RHO join with table sizes 100 (build) and \qty{400}{\mega\byte} (probe). Comparison between Plain CPU, SGX Data in Enclave and Plain CPU with SSB M(itigation) enabled. Applying our unrolling and reordering optimization improves the performance of the slower phases significantly.}
    \label{fig:RHO-stages}
\end{figure}

\paragraph{Finding the root cause.}
To investigate the reason for this slowdown, we again break down the join into its main phases. The upper part of \cref{fig:RHO-stages} compares the runtimes of the stages in a single-threaded RHO join between the plain CPU (blue bar) and the enclave (green). It reveals that the overhead largely originates from creating histograms (\emph{Hist. 1/2}) for radix partitioning and the partitioning itself (\emph{Copy 1/2}). Especially the histogram creation is up to 2 times slower inside the SGX enclave.

This is curious because the base operations required to create a histogram for radix partitioning do not have high overheads. As shown in \cref{sec:scans}, linear main memory reads only have \qty{3}{\percent} overhead in SGXv2 enclaves. Additionally, the previous micro-benchmark revealed that random cache accesses have no overhead inside enclaves (\cref{sec:join-random-access}). Hence, it is unlikely that this overhead is caused by memory encryption.

Indeed, further investigation isolated the issue source to read-dependent write positions. That means all algorithms that alternately read values, determine a write position from the read, and then write to the determined position are affected. Their performance in SGX enclaves is reduced even if they are not bottlenecked by memory accesses. The histogram is such an algorithm because the histogram bin that must be written (incremented) depends on the input key. Other affected algorithms in our experiments are the copy phase of radix partitioning, the hash table build phase of the RHO join, and the hash table build phase of the PHT join discussed in the previous section.

\paragraph{Explaining the slowdown.}
After identifying the slowdown and consultations with Intel, we find that the microcode mitigation against Spectre Version 4, also known as \ac{SSB} \cite{intelcorporationSpeculativeStoreBypass2018}, matches the observed behavior and is a possible explanation. The \ac{SSB} vulnerability arises from the fact that the CPU speculates if a load position overlaps with the currently unknown position of an unfinished preceding store. If the speculation correctly assumes that the store does not invalidate the load, the load can be done in parallel to the store. However, wrong speculations cause reads of stale data. Although the CPU will detect this and discard values calculated from the stale read, this behavior can be exploited to create a side channel, potentially exposing the secrets of an application \cite{intelcorporationSpeculativeStoreBypass2018}. As one option to mitigate this side channel, Intel introduced a microcode patch that exposes a switch to disable the speculative behavior. With this switch activated, loads will not start before the addresses of all preceding stores are known. On a plain CPU, the mitigation is disabled by default. In SGX enclaves, however, the mitigation is permanently enabled and cannot be disabled \cite{intelcorporationSpeculativeStoreBypass2018}.

To verify if this mitigation causes the performance difference between plain CPU and enclave computation, we enabled the mitigation outside of the enclave with the \texttt{prctl} \cite{variousauthorsPrctlLinuxManual2023} function. As the setting Plain CPU M(itigation) in \cref{fig:RHO-stages} shows, enabling the mitigation outside the enclave increases the runtime to exactly the same time as inside the enclave. Thus, the performance difference can be explained fully with the mitigation. We verified that this issue still exists on Emerald Rapids processors. This raises the question if the performance overhead of this side channel mitigation can be counteracted with specific optimizations.

\paragraph{Addressing the slowdown.}
To the best of our knowledge, we are the first to identify these significant performance issues caused by this side channel mitigation and propose a solution.
As explained in the next section, since the side channel mitigation essentially deactivates the speculative execution in enclave mode, we suggest compensating this effect with loop unrolling and instruction reordering, as exemplified in \cref{lst:unrolled} for the histogram computation.
Our experimental results in \cref{fig:RHO-stages} (lower part) confirm that our suggested optimization effectively counteracts the slowdowns caused by the SSB mitigation. Histogram and hash table build performance with the optimization applied are nearly equal in the enclave. For the copy step, the results are more nuanced. First, the optimization improves the performance of the copy step in all three settings. The performance of the plain CPU baseline also increases because the unrolling reduces the frequency of expensive miss-speculations. Second, since the copy operation inherently has dependent loads and stores that cannot easily be split up, the optimization cannot remove all performance differences for this algorithm. Third, there is a remaining performance difference between the enclave and mitigation settings. This can be explained by the random access pattern of copying tuples to their partitions.  All in all, the unroll and reorder optimization decreases the runtime of the single-threaded radix join by \qty{62}{\percent} and increases the relative throughput from \qty{46}{\percent} to \qty{91}{\percent} of the baseline without mitigation and to \qty{97}{\percent} of the baseline with mitigation enabled.

\begin{figure}
    \centering
    \includegraphics{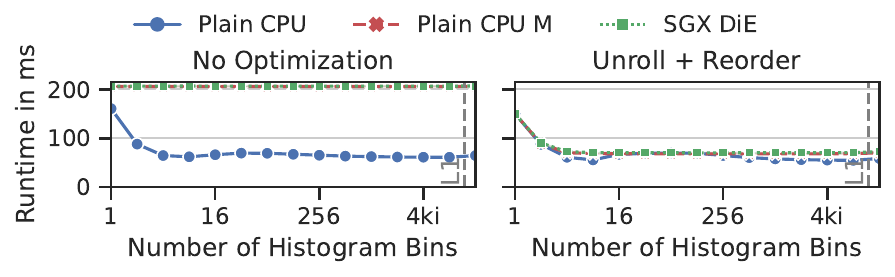}
    \caption{Histogram micro-benchmark for typical numbers of histogram bins. Using the scalar code, histogram creation is \qty{225}{\percent} slower when executed inside the enclave. Manual loop unrolling and instruction reordering decreases the slowdown to less than \qty{20}{\percent}.}
    \label{fig:histogram-benchmark}
\end{figure}

\paragraph{Optimization in depth.}
Since the slowdown is triggered by stores to data-dependent positions followed by loads, it can be reduced by increasing the time between data-dependent stores and following loads. One strategy to achieve this time separation is grouping loads and stores, i.e., first loading multiple values and then dispatching multiple stores in sequence. Thereby, most store positions are determined in parallel to other stores and do not block loads. This increases the number of concurrent memory operations and decreases average latency.
This effect can be achieved by unrolling the inner loop of the algorithm and reordering the instructions. An example of the histogram creation loop is shown in \cref{lst:unrolled}. In every iteration, the algorithm first reads multiple keys from the input and calculates indexes, and then issues multiple increments to the determined indexes in sequence. This optimization decreases the runtime of histogram creation in SGX enclaves to within \qty{20}{\percent} of the same code running in normal CPU mode (cf. micro-benchmark in \cref{fig:histogram-benchmark}).

\begin{listing}
\begin{minted}
[
fontsize=\footnotesize,
]{c++}
// Original histogram loop
for (uint32_t i = 0; i < data_size; ++i) {
    size_t idx = (data[i].key & mask) >> shift;
    ++hist[idx];
}

// Histogram loop with unroll + reorder optimization
uint32_t i = 0;
for (; i + 8 <= data_size; i += 8) {
    size_t idx0 = (data[i].key & mask) >> shift;
    size_t idx1 = (data[i+1].key & mask) >> shift;
    ...
    size_t idx7 = (data[i+7].key & mask) >> shift;
    ++hist[idx0];
    ++hist[idx1];
    ...
    ++hist[idx7];
}
\end{minted}
\caption{First loop: Histogram creation code used in radix partitioning. The table \texttt{data} is scanned, a simple hash function is applied to the join keys, and corresponding histogram bins are incremented. Second loop: Histogram creation for radix partitioning unrolled 8 times (shortened).}
\label{lst:unrolled}
\vspace{1em}
\end{listing}

\begin{figure}
    \centering
    \includegraphics{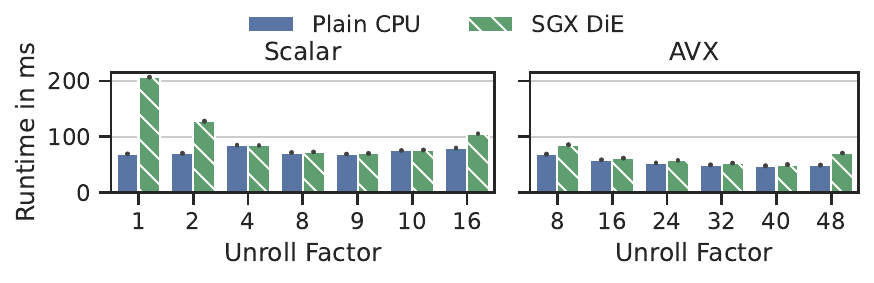}
    \caption{Runtime of the histogram algorithm in an enclave for a \qty{500}{\mega\byte} input array and 32 bins for varying unrolling depth. Left: scalar code. Right: Vectorized code using AVX.}
    \label{fig:histogram-unroll-factors}
\end{figure}

\paragraph{Tuning of the optimization.}
Theoretically, the further the code is unrolled, the better the performance should become, as the average latency per load/store caused by the mitigation decreases. In practice, this improvement is limited by the number of registers available to store write positions. The number of registers available for this purpose depends on the algorithm and the number of required registers for its calculation. As depicted on the left side of \cref{fig:histogram-unroll-factors}, on the Ice Lake CPU architecture, runtime for the histogram algorithm decreases until 9x unrolling, where 9 indexes are determined and incremented per iteration. Starting from 10x unrolling, at least one index must be stored on the stack and loaded again. This interleaves loads and stores, thereby introducing additional latency and decreasing performance. Similarly to the histogram algorithm, the optimal unrolling factor can be determined empirically for other algorithms.

A prominent option to increase the number of indexes that can be stored in registers further is the usage of AVX for index calculation. Thus, we implemented index calculation for histograms with AVX intrinsics and unrolled the loop to store more indexes in registers. As depicted on the right side of \cref{fig:histogram-unroll-factors}, this can improve performance further. Unrolling the loop 5 times and calculating 40 indexes (since each load consumes 8 indexes) before the increments start achieved the best performance in our experiments and a performance difference of \qty{5}{\percent} between the enclave and plain CPU execution. However, when unrolling 6 times or more, the compiler generates instructions that store intermediate registers on the stack, decreasing performance.

\begin{figure}
    \centering
    \includegraphics{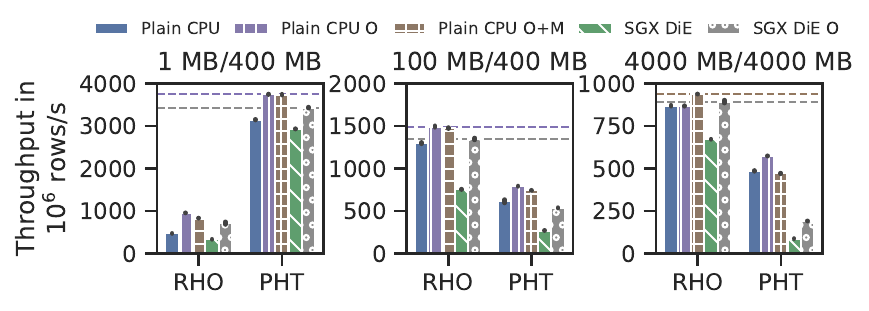}
    \caption{Comparison of RHO and PHT throughput joining three different table sizes with 16 threads with and without optimization (O) and \ac{SSB} mitigation (M). Both joins profit from the optimization. With the optimization, the optimal algorithm for a specific table size reaches more than \qty{90}{\percent} of the fastest baseline throughput (horizontal lines).}
    \label{fig:radix-improvements}
\end{figure}

\paragraph{Putting it all together.}
Finally, we investigate the effect of manual loop unrolling and instruction reordering on RHO and PHT using multi-threaded execution with all 16 cores on one socket.
Again, we compare the join throughput inside the enclave to the same join code executed without SGX protection. We use the optimal unrolling factors for all algorithms marked with O(ptimized). Additionally, we show the effect of changing input table sizes with three example sizes. The results are depicted in \cref{fig:radix-improvements}. With the optimization applied, the RHO join performance inside the enclave improves by \qty{114}{\percent} to \qty{33}{\percent} (SGX DiE compared with SGX DiE O). Thereby, it achieves \qty{75}{\percent} to \qty{95}{\percent} throughput of the fastest RHO plain CPU baseline. RHO performs best in the \qty{100}{\mega\byte}/\qty{400}{\mega\byte} setting because of the relatively equal input table sizes that still fit the L1 TLB during partitioning.
The PHT join throughput improves by \qty{17}{\percent} to \qty{118}{\percent}. For the small build size, it achieves \qty{92}{\percent} performance of the baseline. For the larger build sizes, it reaches only \qty{67}{\percent}/\qty{33}{\percent} performance of Plain CPU O since it is still limited by slower random main memory access. Thus, while the optimal join algorithm depends on the data characteristics, with our optimizations, the performance degradation inside the enclave is less than \qty{10}{\percent}, as visualized by the dashed horizontal lines in \cref{fig:radix-improvements}.

\begin{lessons}[after skip=0pt]
\paragraph{Lessons learned.}
The side channel mitigation for Spectre V4, which is disabled by default outside of enclaves but 
forcibly enabled inside enclaves, increases the runtime of the investigated algorithms by up to \qty{225}{\percent}. All algorithms that determine write positions from input values, such as histogram creation, hash table creation, and partitioning, are affected. However, we show that loop unrolling and instruction reordering, as well as vectorization, can improve the performance of affected algorithms.
\end{lessons}

While previous work noticed the potentially negative effects of microcode updates on performance \cite{vaucherShortPaperStressSGX2019}, we are the first to pinpoint a concrete side channel mitigation as the root cause, show the effect on databases, and suggest solutions.

More generally, the results show that, especially with other overheads like EPC paging removed, side channel mitigations can play a large role in software performance for SGX enclaves. Thus, developers should be aware of them and consider optimizations to circumvent performance regressions. Since the slowdown caused by this mitigation can be large, we suggest mentioning it and other mitigations applied in enclaves in the Intel SGX Developer Guide, which already contains other advice for performance in SGX enclaves~\cite{intelcorporationIntelSoftwareGuard2023}.

\subsection{Analyzing NUMA Effects for Joins}
\label{sec:join-numa}

As introduced in \cref{sec:background}, a new feature of SGXv2 is the support for servers with multiple sockets, and enclaves leveraging the secure memory on multiple NUMA nodes.
Communication between NUMA nodes is known as an important performance factor for in-memory database operations and, in particular, joins~\cite{faerberMainMemoryDatabase2017,kieferExperimentalEvaluationNUMA2013}. Moreover, while several NUMA optimizations exist to increase NUMA locality, cross-NUMA traffic cannot be prevented, particularly for complex queries (e.g., those including multiple joins). 
Hence, in this section, we aim to analyze the effects of cross-NUMA traffic.
Since enclave communication via the UPI is encrypted~\cite{johnsonSupportingIntelSGX2021} and previous work measured an increase in latency when accessing memory across NUMA boundaries in SGX compared to accessing cross-NUMA without SGX~\cite{el-hindiBenchmarkingSecondGeneration2022}, we investigate how these costs influence the performance of join algorithms. An investigation of encrypted UPI throughput is contained in \cref{sec:scans-numa}.

The main issue of NUMA in the current SGXv2 is the fact that the main tools for NUMA optimization -- memory allocation and thread pinning in specific NUMA regions -- are features of the untrusted OS and not available in enclaves. Thus, it is impossible to ensure local processing in the general case. For the following experiment, we make use of the fact that the Linux kernel on our trusted benchmark machine allocates EPC pages of an enclave in the local NUMA region if possible.

\paragraph{Benchmarking extreme NUMA cases.} Since optimizations for NUMA in joins is a wide research field on its own, we concentrate on extreme cases in our experiments and expect the performance of real-world cases to fall in between. Our optimal baselines are a NUMA-local join with 16 threads (Single Socket Plain CPU) and a join where both input tables are pre-partitioned on the join key to both NUMA nodes (Dual Socket Plain CPU). The second setup avoids cross-NUMA traffic completely and reaches double throughput of only using one NUMA region. 
Additionally, we analyze two cases in SGX. The first is a NUMA-local join, where the enclave and all its memory are located on NUMA node 0 and the join is executed by all threads on the same node (Single Socket SGX DiE). In the second setting, all 32 cores of both sockets in the system execute the join, but the enclave and all its memory are allocated exclusively on one of the nodes (Dual Socket SGX DiE).

\begin{figure}
    \centering
    \includegraphics{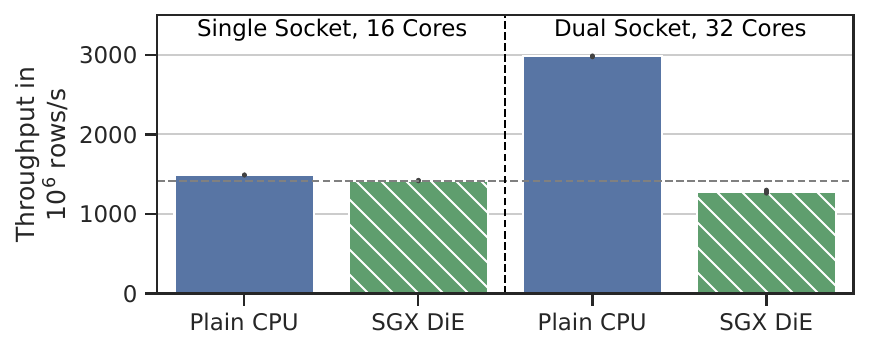}
    \caption{Throughput of an RHO join on a NUMA system in worst and best cases. A partitioned join with double the cores can achieve double throughput on a plain CPU. Inside the SGXv2 enclave, adding a second socket with 16 cores reduced the join performance because of the remote data access.}
    \label{fig:join-numa}
\end{figure}

The results in \cref{fig:join-numa} show that without manual intervention, the use of cores from multiple CPU sockets can decrease the performance of a join executed inside SGX enclaves instead of increasing it as intended.
By comparing SGX Data in Enclave (DiE) on a single socket with 16 cores to SGX DiE running on both sockets with 32 cores, it is clear that adding another 16 threads to the join while data is not distributed over both nodes decreases the join throughput inside the enclave instead of increasing it. This wastes the CPU cycles of 16 cores. Thus, the SGX join with 32 cores achieves less than half of the optimal case performance for a join that leverages all cores (Dual Socket Plain CPU).

\begin{lessons}
\paragraph{Lessons learned.} Cross-NUMA memory access significantly reduces the performance of joins in SGXv2 enclaves. To improve this situation, NUMA-aware memory allocations and thread placement are required. However, since the OS manages these hardware features, such manual control could currently only be implemented when trusting the OS to do thread pinning and memory allocations on specific CPUs correctly. As such, depending on the setting, NUMA-awareness can not be achieved in SGXv2.
\end{lessons}

\subsection{Synchronization \& Memory Allocation}
\label{sec:join-sdk}

To conclude the investigation of join performance, we discuss the last remaining factors in our \emph{Mixed effects} category of SGXv2 overheads: Slowdowns caused by SGX SDK mutexes and dynamic enclave memory allocation.

\paragraph{Effects of mutexes.}
Many multi-threaded join implementations require synchronization of threads during execution. The authors of TEEBench~\cite{maliszewskiWhatPriceJoining2021} showed that the SGX SDK mutex limited the join performance in SGXv1 because it causes costly context switches outside the enclave. We revisit this issue in the context of \sgxTwo{} since efficient synchronization becomes more important due to the new hardware: The increased number of hardware threads that can create more contention and other bottlenecks like EPC size have been removed.

\begin{figure}
    \centering
    \includegraphics{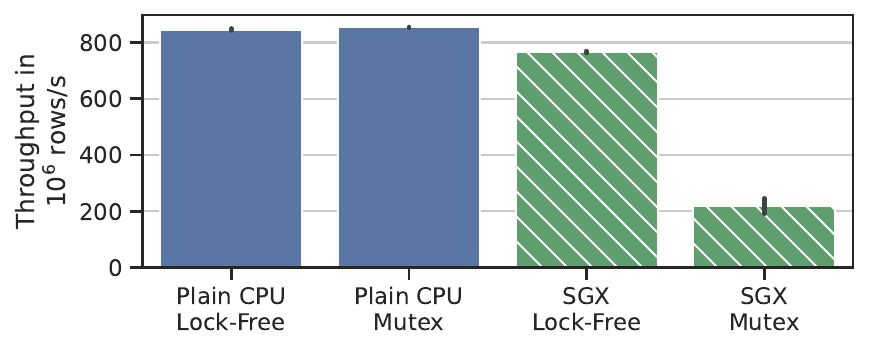}
    \caption{Throughput of an RHO join with contention on the task queue. Outside of the SGX enclave, the queue choice does not make a significant difference. Inside the enclave, protecting the queue with a mutex instead of a lock-free design reduces the throughput by \qty{75}{\percent}.}
    \label{fig:join-mutex-contention}
\end{figure}

To investigate the overhead in SGXv2, we designed the following experiment: We switched out the lock-free task queue of our radix join, distributing partition and join tasks between cores, with the mutex-guarded queue used in the original TEEBench. Since the issue only occurs in case of contention, we forced contention on the mutex by using small join partitions and 16 threads. In \cref{fig:join-mutex-contention}, we compare the performance of both implementations inside an SGX enclave and on a plain CPU.

The experiment results exemplify that replacing high-overhead SDK functions with more optimized solutions can dramatically change the performance characteristics of an algorithm in SGXv2. Outside of the enclave, the choice of queue implementation does not cause significant throughput differences (blue bars). However, inside the SGX enclave, join throughput drops by \qty{75}{\percent} when comparing the lock-free queue that avoids OS interactions with the mutex-guarded queue (green bars).

\paragraph{Root cause of mutex slowdowns.}
The observed performance difference is caused by the SGX SDK mutex' design. In this design, threads transition out of the enclave to sleep when they encounter a locked mutex. When the owning thread unlocks the mutex, it exits the enclave to wake the first waiting thread up. Then, both re-enter the enclave. During these enclave transitions, the mutex stays locked, extending the critical section and increasing the probability of threads arriving at a locked mutex.
This is sensible if the critical section protected by the mutex is significantly longer than an enclave transition. However, critical sections of in-memory join algorithms are orders of magnitude shorter than enclave transitions. Thus, a mutex-based design has a negative performance impact.
For the joins in this paper, we considered this effect and replaced mutexes found in their implementations with spin locks or lock-free data structures.

\paragraph{Effects of memory allocation.} \label{par:EDMM}
Memory management is another critical performance factor for DBMSs \cite{durnerImpactMemoryAllocation2019}. Therefore, many real-world databases use buffer managers that pre-allocate memory before it is needed. In cloud settings, however, it is desirable not to pre-allocate all memory available to a server at start time \cite{aroraFlexibleResourceAllocation2023}. Additionally, before a query is started, it is not always clear how much memory the execution and result materialization will require. Therefore, DBMSs can be forced to allocate additional memory dynamically during query execution. 

Originally, SGX enclaves had a fixed size that could not be changed after enclave creation. With the SGX 2 instruction set, Intel introduced additional CPU instructions that enable securely adding and removing enclave pages at runtime \cite{mckeenIntelSoftwareGuard2016}. This feature is called \ac{EDMM} and available since Linux version 6.0/SGX SDK version 2.18 \cite{intelcorporationIntelSoftwareGuard2022}. Depending on the enclave settings, \ac{EDMM} is either transparent to the developer or must be managed explicitly.
The following experiment investigates the performance implications of automatic \ac{EDMM}.
In this experiment, we run our \sgxTwo-optimized RHO join and additionally materialize the result table. By reducing the amount of pre-allocated memory during enclave start to a minimum, we force a situation where all memory required to write the join result tuples must be allocated by dynamically increasing the enclave size. We compare this to a setting where the enclave is large enough to fit all result tuples without adding memory (static enclave size). As additional baselines, we also execute the join in native mode with memory pre-allocated (pre-alloc) and with dynamic memory allocation (dynamic alloc.).

\begin{figure}
    \centering
    \includegraphics{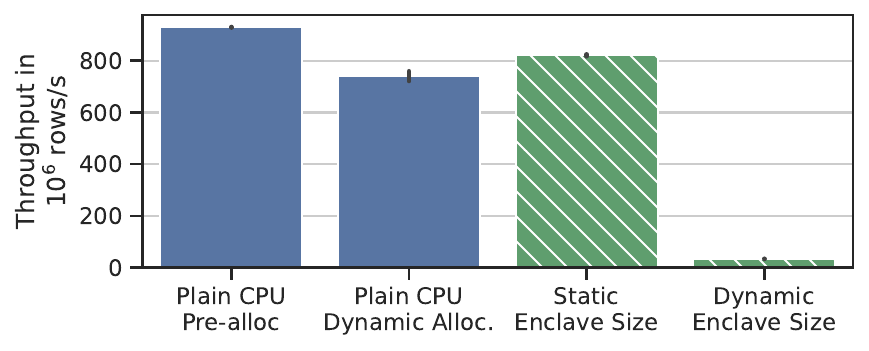}
    \caption{Throughput of the RHO join materializing output tuples inside a statically sized pre-allocated enclave compared with the throughput of the same join in a dynamically sized enclave. Dynamically increasing the enclave size during the join reduces its performance by \qty{95}{\percent}.}
    \label{fig:join-dynamic}
\end{figure}

The results in \cref{fig:join-dynamic} show that dynamically increasing the enclave size for memory allocations is an order of magnitude more expensive than dynamic memory management outside enclaves.
In the experiment, the join inside the enclave achieves only \qty{4.6}{\percent} throughput compared to the plain CPU join that allocates dynamically and only \qty{4.1}{\percent} throughput compared to the static enclave size setup. The difference can be attributed to the security protocols necessary for resizing enclaves \cite{mckeenIntelSoftwareGuard2016}.

\begin{lessons}
\paragraph{Lessons learned.} The SGX SDK mutex and EDMM can introduce large overheads for query execution in enclaves. Thus, lock-free data structures should be preferred in enclaves, and EDMM should either not be used or actively managed to prevent its overheads.
\end{lessons}

\section{Scans in SGXv2}
\label{sec:scans}

In addition to joins, table scans are essential for the performance of OLAP systems since they require scanning large amounts of data with very high throughput.
In this section, we use a columnar SIMD scan as a typical scan algorithm in OLAP databases which causes high demands on the memory subsystem.

\paragraph{Importance of the analysis.}
Interestingly, previous work incorrectly reported the lack of SIMD instructions inside SGX enclaves~\cite{maliszewskiWhatPriceJoining2021,maliszewskiCrackingLikeJoinTrusted2023} and hence state-of-the-art vectorized scan algorithms~\cite{willhalmSIMDscanUltraFast2009, polychroniouRethinkingSIMDVectorization2015} have not been studied yet.
Given the high core counts available in recent server processors and the new memory encryption technology used in SGXv2~\cite{johnsonSupportingIntelSGX2021}, it is unclear if the memory decryption engine is fast enough to allow for high throughput scans with multiple cores. 
Similarly, the impact of the additional encryption on the scan throughput when crossing NUMA boundaries has not been explored yet.
Thus, studying throughput-optimized column scans is essential for understanding the performance characteristics of \sgxTwo{} for DBMSs.

\paragraph{Scan algorithm and data.}
For our benchmarks, we implemented state-of-the-art scan algorithms~\cite{willhalmSIMDscanUltraFast2009,polychroniouRethinkingSIMDVectorization2015} using AVX\,512 instructions.
Our implementations load 64 byte-sized values at once from a column, compare them to a lower and upper bound (i.e., incorporating a filter condition), and store the comparison result either in a bit vector or, as we show in a later experiment, materialize row identifiers.
We use the byte-aligned input value size for this experiment to minimize the required processing effort and maximize the throughput requirements on the memory system.
As in our join benchmarks, we assume that the memory for the scan result is pre-allocated to see the pure overhead of memory encryption and decryption. We determine the scan throughput by dividing the size of the compressed input array by the scan runtime.

\begin{figure}
	\centering
	\includegraphics[width=\linewidth]{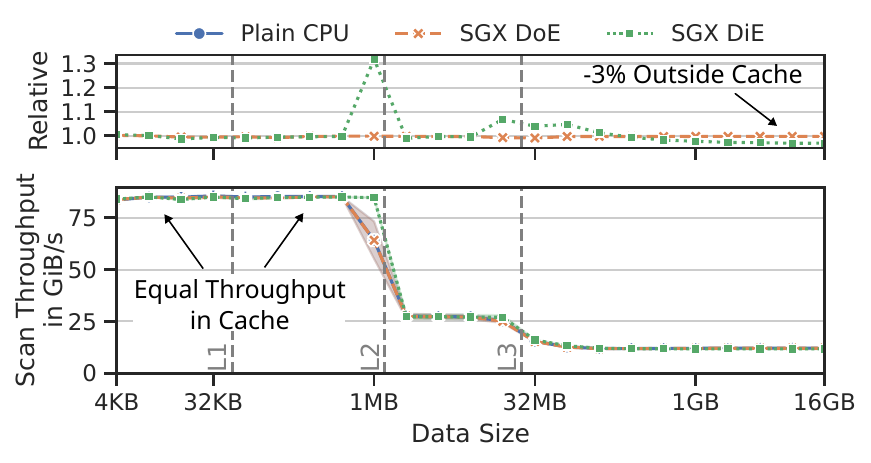}
	\caption{Read throughput of a scan using AVX\,512 instructions, scanning over the same data 1,000 times. Comparison between enclave code reading enclave data (DiE), enclave code reading unencrypted data, and non-enclave code reading unencrypted data (plain). Inside the cache, scan throughput is equal, outside the cache we observe a slowdown of \qty{3}{\percent}.}
	\label{fig:simd-scan-cache}
\end{figure}

\subsection{Single-Threaded Column Scans}
Before stressing the limits of the memory encryption engine using multiple threads, we start by analyzing the encryption/\allowbreak{}decryption overhead for a single-threaded scan.
To this end, we compare the read throughput of a column scan on a single CPU core between enclave code reading enclave data (SGX Data in Enclave), enclave code reading plain data (SGX Data outside Enclave), and our baseline, non-enclave code reading plain data (Plain CPU). Additionally, we vary the size of the scanned column from \qty{4}{\kilo\byte} to \qty{16}{\giga\byte}.
To show the effect of CPU caches, we first execute 10 warm-up scans and afterward start the time measurement for another 1000 scans.

As we see on the left side of~\Cref{fig:simd-scan-cache}, again, there is no SGX-inherent overhead if data is cache resident. This is expected because data in caches is in plain text and does not require any decryption.
When the data does not fit into the L3 cache, the column scan over encrypted enclave data (stored in the EPC) is only minimally (i.e., $\approx$3\,\%) slower than the scan over unencrypted data.
This is a clear improvement over SGXv1, which showed a much larger performance loss of up to \qty{75}{\percent} even for simple non-vectorized scans~\cite{maliszewskiCrackingLikeJoinTrusted2023}.

\subsection{Multi-threaded Execution}
\label{sec:good-multithreading}

\begin{figure}
	\centering
	\includegraphics[width=\linewidth]{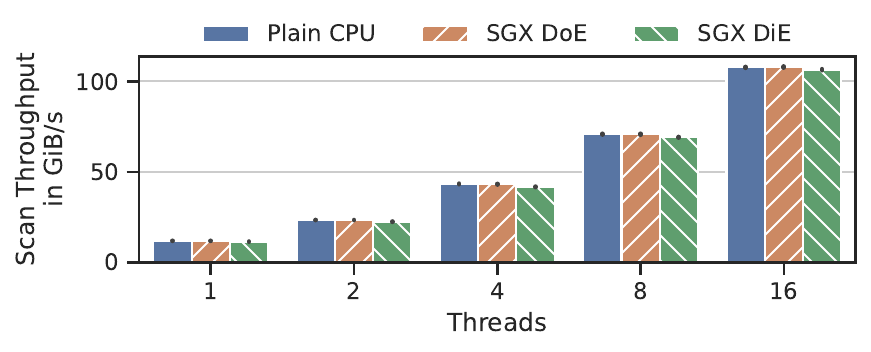}
	\caption{Column scan throughput scales with more threads. Scaling behavior is equal between running inside the enclave and outside. There seems to be no bottleneck caused by memory encryption or decryption in \sgxTwo.}
	\label{fig:simd-scan-multithreading}
\end{figure}

Next, we explore if the memory encryption engine inside SGXv2 becomes a bottleneck when increasing the scan throughput by using multi-threading, as it is done in many modern \ac{DBMS}.

To do this, we execute the same scan algorithm as in the previous experiment over a \qty{16}{\giga\byte} column while scaling the number of used cores from 1 to 16.
As shown in \cref{fig:simd-scan-multithreading}, the enclave memory protection mechanisms do not become a bottleneck on our processor. The scaling behavior is equal between SGX and plain CPU. Further, in both settings, our algorithm is able to reach the memory bandwidth limit with 16 cores. We verified this with Intel VTune for the plain CPU scan.

\subsection{Scans with Varying Read/Write Ratio}
\label{sec:good-write-ratio}

The previous experiments are both read-heavy and only have to write a small output as tightly packed bit vectors. 
As a consequence, the memory encryption engine mainly performs decryption when loading data from the EPC and only a limited amount of encryption.
This leaves open the question if increased amounts of writes stress the memory encryption to a degree where it cannot keep up with the column scan. To check if the ratio of reads and writes to memory influences the performance of scans inside the enclave, we evaluate a second scan with a variable write ratio (i.e., by using different selectivities). Instead of a bit vector, the second scan implementation returns 64-bit integers (i.e., row indexes) for the values that match the range criterion. Since a 64-bit index is 8 times larger than an 8-bit value, the write rate of this scan is 8 times the selectivity.

\begin{figure}
    \centering
    \includegraphics[width=\linewidth]{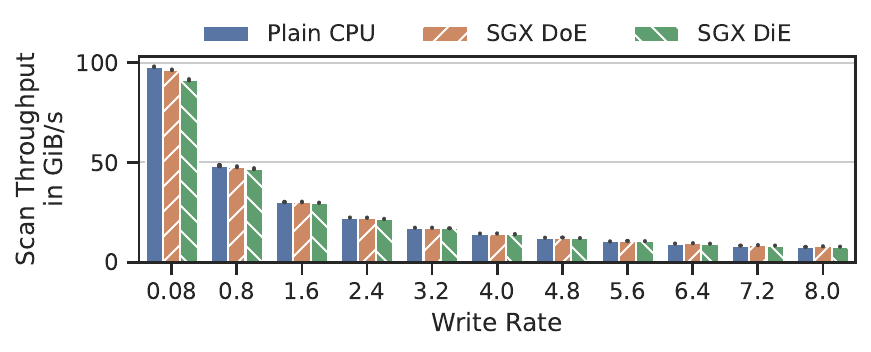}
    \caption{Varying selectivity to increase the write rate of the scan benchmark. Uses a scan that returns matching indexes. Size of input: \qty{4}{\giga\byte}. 16 Threads. An increased write rate does not cause an increased overhead in \sgxTwo.}
    \label{fig:simd-scan-write-rate}
\end{figure}

As can be seen in \cref{fig:simd-scan-write-rate}, an increased write rate does not lead to a higher reduction of the read throughput inside the enclave compared to outside. The read throughput of the column scan decreases to the same degree inside and outside the enclave.

\begin{lessons}
\paragraph{Lessons learned.} The SGXv2 memory encryption mechanism causes minor overheads for column scans optimized for maximum memory throughput.
The performance of this operation is, generally speaking, equivalent between normal CPU and enclave mode. 
This insight is independent of the number of CPUs employed for the scan and the ratio of reads and writes. 
We expect that other bandwidth-bound algorithms with linear access patterns, such as scalar functions, will behave similarly.
\end{lessons}

\subsection{Scans and NUMA}
\label{sec:scans-numa}

As introduced in \cref{sec:background,sec:join-numa}, \sgxTwo{} supports enclaves on multi-socket servers. In the context of scans, this theoretically enables the utilization of additional cores available on the second NUMA node, further parallelizing scan algorithms to increase performance. However, as NUMA-local memory allocations and thread pinning are currently not available in \sgxTwo{} enclaves, scan threads may be forced to access EPC data on remote nodes over the \ac{UPI} link, which incurs additional overhead due to encryption.
To quantify the overhead of UPI encryption, we analyze the throughput characteristics of cross-NUMA scans.

We again benchmark extreme cases and use the observation that the Linux kernel allocates EPC pages on the local node. To build a cross-NUMA column scan benchmark, we pin the scan execution threads to the node on which the enclave was not allocated.
This ensures that all read and write operations cross the UPI link.
Using this technique, we compare the read throughput of a NUMA-local plain CPU scan with the performance of a cross-NUMA plain CPU scan and a cross-NUMA scan reading and writing encrypted data inside an SGXv2 enclave. The benchmarked scans use 1 to 16 threads.

\begin{figure}
    \centering
    \includegraphics[width=\linewidth]{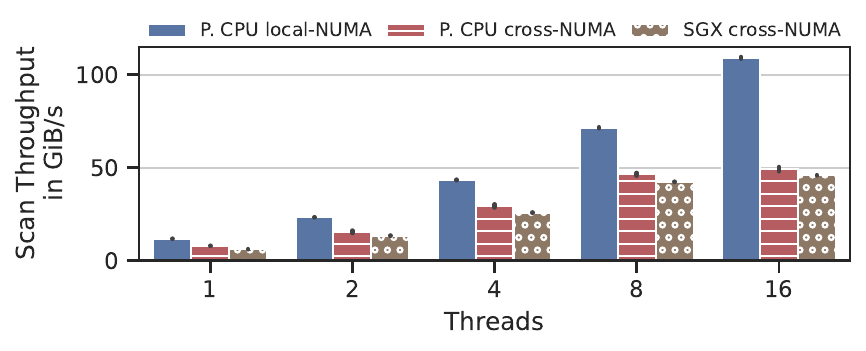}
    \caption{Cross-NUMA column scan throughput in an SGX enclave compared with a cross-NUMA scan without SGX and a local-NUMA scan without SGX. UPI traffic encryption in SGX causes an additional performance decrease.}
    \label{fig:simd-scan-cross-numa}
\end{figure}

\Cref{fig:simd-scan-cross-numa} shows the results of our benchmark. The measurements show a lower throughput for cross-NUMA scans, especially when using multiple threads. It is important to note here, that the theoretical upper bound for throughput of the 3 UPI links between the sockets in our server is \qty{67.2}{\giga\byte\per\second} and executing the scan with 8 and 16 threads approaches this upper limit. When comparing the plain CPU cross-NUMA scan performance with its enclave counterpart, we measured \qty{77}{\percent} of the baseline throughput with a single thread. This relative performance increases with the number of threads up to \qty{96}{\percent} for 16 threads, where the scan is bound by the general speed of the UPI links.

\begin{lessons}[after skip=0pt]
\paragraph{Lessons learned.} Cross-NUMA memory access reduces scan throughput in SGX enclaves further than outside enclaves. Thus, optimizations for local access would have a more significant effect if they were possible.
\end{lessons}

\begin{figure}
    \centering
    \includegraphics{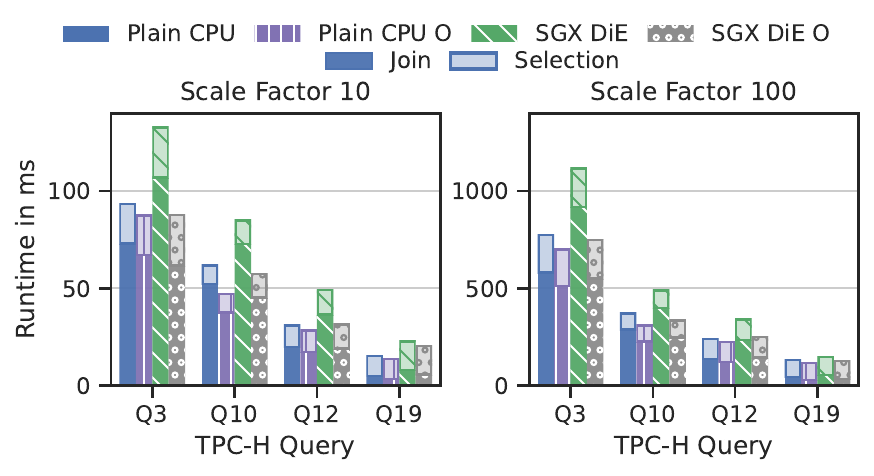}
    \caption{Runtime of four TPC-H queries at scale factor 10 and 100 using the RHO join. Comparison between outside the enclave and inside the enclave, both with and without optimization. Our optimization reduces the performance difference between outside and inside the enclave.}
    \label{fig:full-query}
\end{figure}

\section{Composing Operators in Queries}
\label{sec:full-queries}

Finally, we investigate the performance of our optimized join and scan operators when composed in query plans.
The goals of this experiment are threefold. First, we examine if the effect of the unrolling and instruction reordering optimization is still relevant in the bigger picture. Second, we investigate the influence of result materialization and TPC-H data characteristics (different table sizes, wider SIMD-scan input values, selective joins, and different payload data) on performance. Third, we assess if the overall query execution performance in an SGXv2 enclave is competitive with the native setting.

For this evaluation, we used TPC-H queries 3, 10, 12, and 19 as workload because these queries mainly consist of scans and joins. We run the queries with the TPC-H data at scale factor 10 and 100 as input. To see the effects of the operators investigated in this paper more clearly, we remove all other operators, replace the final aggregation with count(*), and represent dates and categorical strings as integers, mimicking the evaluation setup for CrkJoin~\cite{maliszewskiCrackingLikeJoinTrusted2023}. All operators and queries are implemented in our C++ framework and compiled before execution. The queries are implemented using the optimized RHO join from \cref{sec:join}. All 16 cores available on one hardware socket are used for parallelism.

The results in \cref{fig:full-query} show that the optimizations introduced in the previous sections (settings annotated with O) indeed result in performance improvements on the query level and reduce the query runtime by \qty{12}{\percent} (Q19, SF10) to \qty{39}{\percent} (Q12, SF10) compared to the unoptimized version. Compared to the execution on the native CPU, the overhead of running the queries in SGX enclaves is reduced from \qty{38}{\percent} on average to \qty{14}{\percent}. As expected, scan \& selection performance is very similar across settings. Therefore, the performance difference between the enclave and native setup primarily originates from the join implementation. Result materialization and the data characteristics of TPC-H do not introduce unexpected performance differences between native and SGXv2 execution.

\begin{lessons}
\paragraph{Lessons learned.}
Using state-of-the-art operator implementations combined with SGXv2-specific optimizations enables query plan execution at near-native performance inside an enclave.
\end{lessons}

\section{Discussion of a Performance Model}
\label{sec:summary}

Given the evaluation results of our paper for in-memory query operators, one remaining question is: How can the findings be transferred to other algorithms and data structures? For such a task, a unified cost model that incorporates memory access costs and optimizations for query execution operators is required. 

As a first approximation of such a model, we summarized our findings in 
\cref{tab:model}. Given the memory access pattern, as well as the processed data size, the table shows the estimated slowdown to be expected of an algorithm running in an SGXv2 enclave compared to the plain CPU. This table can already guide performance optimizations, such as partitioning or manual unrolling and reordering (as presented in our paper). For example, as we have seen in \cref{sec:join-pipelining}, histogram computations have a data-dependent write memory access pattern. Thus, even if data fits in the last level cache (LLC), we can expect a high overhead of up to \qty{225}{\percent} and thus should consider applying additional optimization in an SGXv2 enclave.

Developing a more sophisticated model that can capture more nuanced performance aspects, such as the order of read and write accesses, and predict exact algorithm performance or slowdown is very involved, as it requires modeling interactions between access patterns, concrete implementation, and CPU optimizations like speculative execution. For example, Manegold et al. \cite{manegoldGenericDatabaseCost2002} created a performance model for in-memory query operators for traditional single-core CPUs, and we believe that such a cost model would be a good starting point. Still, it needs to be updated to reflect not only the effects of multi-cores and NUMA but also SGX-related aspects, such as the effect of data-dependent writes, which we have seen to severely impact the performance of query operators. As such, creating such a more detailed model is out of the scope of this paper but represents an important avenue of future work.

\begin{table}
    \fontsize{8}{8.1}
    \centering
    \caption{High-level performance model for the expected slowdown of algorithms in SGXv2 enclaves. As shown in our work, performance is mainly influenced by data size (DS) and the access pattern (AP) of an algorithm.}
    \begin{tabular}{|c|c|c|c|}
        \hline
         \diagbox[innerleftsep=.09cm,innerrightsep=.09cm]{DS}{AP} & \thead{linear\\read/write} & \thead{independent\\random read/write} & \thead{data-dependent\\write} \\
        \hline
        $< LLC$ & None (\qty{0}{\percent}) & None (\qty{0}{\percent}) & High (up to \qty{225}{\percent}) \\
        $> LLC$  & None (\qty{3}{\percent}) & High (up to \qty{200}{\percent}) & High (up to \qty{800}{\percent}) \\
        \hline
    \end{tabular}
    \label{tab:model}
\end{table}

\section{Related Work}
\label{sec:related}

This study has three main areas of related work: Benchmarks and performance evaluations for SGX, specialized enclave database systems and architectures, and recent evaluations of SGXv2.

\paragraph{Benchmarks and performance evaluations for SGX.} Multiple papers introduce benchmarks suites for 
SGXv1~\cite{vaucherShortPaperStressSGX2019,kumarSGXGaugeComprehensiveBenchmark2022,mahhoukSGXoMeterOpenModular2021} to analyze the performance characteristics of enclaves. Their approach is similar to ours in that they port existing workloads~\cite{kumarSGXGaugeComprehensiveBenchmark2022,mahhoukSGXoMeterOpenModular2021} or benchmark suites \cite{vaucherShortPaperStressSGX2019} to Intel SGX and compare the performance to native execution.
These efforts do not concentrate on specific application domains like databases, and they were conducted before the introduction of SGXv2.

\paragraph{Specialized DBMS for SGX.}
There are multiple proposals for data management systems inside SGXv1 enclaves that suggest approaches to circumvent the performance degradations caused by the limited EPC size \cite{priebeEnclaveDBSecureDatabase2018,antonopoulosAzureSQLDatabase2020,kimShieldStoreShieldedMemory2019,sunBuildingEnclavenativeStorage2021} or investigate the theoretical enclave performance without any memory limit~\cite{vinayagamurthyStealthDBScalableEncrypted2019}.
Most related to our work are the publications by Maliszewski et al.~\cite{maliszewskiWhatPriceJoining2021,maliszewskiCrackingLikeJoinTrusted2023} analyzing the performance of join algorithms in SGXv1 enclaves. They observe that radix joins have beneficial properties for enclaves, but all joins greatly suffer from slow random access and EPC paging. To circumvent these problems, the authors develop CrkJoin~\cite{maliszewskiCrackingLikeJoinTrusted2023} that reaches superior in-enclave performance in their evaluation.
However, our study shows that the CrkJoin optimizations are irrelevant in SGXv2 due to the eliminated EPC bottleneck.
To achieve near-native performance for database workloads in the latest SGX generation, new optimizations and a thorough understanding of the performance characteristics of SGXv2 are required.

\paragraph{SGXv2 performance.} 
To the best of our knowledge, there is still minimal research on the performance characteristics of SGXv2 \cite{el-hindiBenchmarkingSecondGeneration2022,lutschBenchmarkingSecondGeneration2023,miwaAnalyzingPerformanceImpact2023,battistonDuckDBSGX2GoodBad2024b}.
Aside from our previous studies on OLTP workloads \cite{el-hindiBenchmarkingSecondGeneration2022} and neural network inference \cite{lutschBenchmarkingSecondGeneration2023}, Miwa and Matsuo have examined SGXv2's performance for HPC~\cite{miwaAnalyzingPerformanceImpact2023}.
Additionally, Battiston et al. are concurrently studying the performance of running DuckDB inside an SGXv2 enclave~\cite{battistonDuckDBSGX2GoodBad2024b} using the Gramine library operating system~\cite{gramineprojectGramineLibraryOS2023}.
This paper extends our previous work by focusing on modern query execution algorithms, data throughput, and an in-depth investigation of slowdown sources at an architectural level.
Through a detailed analysis of SGXv2 performance characteristics in the OLAP context, we identify new optimizations, such as manual loop unrolling and instruction reordering, 
to enhance the throughput of in-memory algorithms.

\section{Conclusion}
\label{sec:conclusion}

This paper provides a comprehensive analysis of Intel \sgxTwo{}, evaluating its advantages and limitations for secure, high-perfor\-mance analytical databases.
Among other insights, our study offers three main contributions: 
Firstly, we demonstrated that state-of-the-art main memory and cache-optimized join algorithms perform better inside SGXv2 than those optimized for the discontinued SGXv1 due to changed hardware characteristics.
Secondly, we uncovered previously unknown overheads caused by a side channel mitigation enabled inside the enclave but not outside, and we showed how existing algorithms can be optimized to circumvent this slowdown.
Finally, we verified that SGXv2-optimized operators enable the execution of query plans with performance on par with execution outside the enclave.
Overall, our findings highlight the potential of SGXv2 for analytical workloads and show that a deep understanding of its performance characteristics is crucial for designing high-performance DBMSs.

\begin{acks}
This work was supported by the safeFBDC Research Fund of the German Federal Ministry for Economic Affairs and Climate Action (Grant Agreement No 01MK21002K), the National Research Center for Applied Cybersecurity ATHENE, and by SAP SE. We also thank Nicolae Popovici from Intel for reproducing some experiments on the newest processor generation.
\end{acks}

\bibliographystyle{ACM-Reference-Format}
\bibliography{references}

\end{document}